\documentstyle[11pt,aaspp4]{article}

\newcommand\MSUN{M$_{\odot}$}

\newcommand\etal{{\it et al}. }
\newcommand\eg{{\it e. g}. }
\newcommand\ie{{\it i. e}. }
\newcommand\be {\begin{equation}}
\newcommand\en{\end{equation}}

\begin{document}

\title{THE COLLISIONS OF HVCs WITH A MAGNETIZED GASEOUS GALACTIC 
DISK\altaffilmark{4}}     

\author{Alfredo Santill\'an\altaffilmark{1,2}, Jos\'e Franco\altaffilmark{2}, Marco Martos \altaffilmark{2} \and \ Jongsoo Kim\altaffilmark{3}}

\altaffiltext{1}{C\'omputo Aplicado--DGSCA, Universidad Nacional Aut\'onoma de
M\'exico, A. P. 70-264, 04510 M\'exico D.F., M\'exico; Electronic mail: 
alfredo@astroscu.unam.mx}
\altaffiltext{2}{Instituto de Astronom\'\i a, Universidad Nacional Aut\'onoma 
de M\'exico, A. P. 70-264, 04510 M\'exico D.F., M\'exico; Electronic mail: 
pepe@astroscu.unam.mx, marco@astroscu.unam.mx}
\altaffiltext{3}{Korea Astronomy Observatory, San 36-1, Hwaam-Dong,
Yusong-Ku, Taejon 305-348, Korea; Electronic mail: jskim@hanul.issa.re.kr}
\altaffiltext{4}{To appear in {\it The Astrophysical Journal}, the April 20, 
1999 issue, Vol. 515}
                             
\begin{abstract}  
We present two-dimensional MHD numerical simulations for the interaction of
high-velocity clouds with both magnetic and non-magnetic Galactic thick 
gaseous disks. For the magnetic models, the initial magnetic field is oriented 
parallel to the disk, and we consider two different field topologies (with and 
without tension effects): parallel and perpendicular to the plane of motion of 
the clouds. The impinging clouds move in oblique trajectories and fall toward 
the central disk with different initial velocities. The $B$-field lines are 
distorted and compressed during the collision, increasing the field pressure 
and tension. This prevents the cloud material from penetrating into the disk, 
and can even transform a high-velocity inflow into an outflow, moving away from 
the disk. The perturbation creates a complex, turbulent, pattern of MHD waves 
that are able to traverse the disk of the Galaxy, and induce oscillations on 
both sides of the plane. Thus, the magnetic field efficiently transmits the 
perturbation over a large volume, but also acts like a shield that inhibits the 
mass exchange between the halo and the disk. For the non-magnetized cases, we 
also uncover some novel features: the evolution of the shocked layer generates 
a tail that oscillates, creating vorticity and turbulent flows along its 
trajectory.
\end{abstract}
  

\vfill\eject
\section{INTRODUCTION}

High velocity clouds (HVC) are atomic HI cloud complexes located at high
galactic latitudes, and moving with large velocities ($\mid V_{LSR} \mid \geq
90$ km/s) that do not match a simple model of circular rotation for our Galaxy 
(see Mirabel 1981a, Bajaja \etal 1985, and Wakker \& van Woerden 1997). The 
present data indicates an excess of negative-velocity (infall) HVC over 
positive-velocity HVC, but the interpretation for their origin and evolution 
is unclear because their distances and tangential motions are unknown. Limits 
to the location of some particular clouds indicate $z$-heigths of a few 
kiloparsecs (a recent detection of highly ionized HVCs indicates even larger 
distances, but the relationship between ionized and neutral HVCs is unclear; 
Sembach \etal 1998), setting a possible mass range of 10$^{5}$-10$^{6}$ \MSUN 
\ for some of these complexes. Thus, a HVC complex moving with a speed of 100 
km s$^{-1}$ has a kinetic energy of about 10$^{52-53}$ erg. This range of 
values (equivalent to that from an OB association in the disk) indicate that 
the bulk motion of the HVC system, and its interaction with the galactic disk, 
could represent a rich source of energy and momentum for the interstellar 
medium (ISM).

Observational signatures for the interactions of HVCs with galactic disks have 
been claimed in our Galaxy and in some external galaxies. The best known 
examples are in complexes AC and H, both located near the anticenter region 
(Mirabel 1981b; Mirabel \& Morras 1990; Tamanaha 1997; Morras \etal 1998), in 
the direction of the Draco Nebula (Kalberla \etal 1984; Hirth \etal 1985; 
Herbstmeier \etal 1996), in M101 (van der Hulst \& Sancisi 1988), and in NGC 
4631 (Rand \& Stone 1996). 
These observations support the idea that HVC-galaxy collisions can have a
significant influence on the structure and energetics of the gaseous disk.
Previous 2-D and 3-D numerical simulations for collisions with $non-magnetic$ 
(or weakly magnetic) and $thin$ disks (\eg Tenorio-Tagle \etal 1986,1987; 
Franco \etal 1988; Comer\'on \& Torra 1992; Lepine \& Duvert 1994; Rand \& 
Stone 1996), indicate that the resulting ISM structures have sizes of several 
hundreds of parsecs, similar to those ascribed to multi-supernova remnants, 
or superbubbles, from OB associations. The colliding HVCs can sweep a great
amount of disk mass in these non-magnetic models. The resulting shocked layer 
collects both the disk mass and the mass from the HVC, inducing the
formation of massive gas structures far from the midplane (Franco 1986; Alfaro 
\etal 1991; Cabrera-Ca\~no \etal 1996). Moreover, the fastest moving clouds can 
even drill holes through the whole gaseous disk, venting gas into the other 
side of the disk. Thus, for these thin disk cases, an efficient mass 
exchange can result from the interaction of the HVC system with the disk.

The HVC-disk interactions, however, can have a radically different outcome in 
a disk that is thicker and more magnetized than assumed by these previous works 
(see Cox 1990 and Franco \etal 1995). The transition between the main gaseous 
disk and the halo is very broad and complex, with an intricate magnetic field 
structure and a number of extended gas components, including dusty high-$z$ 
layers and the so-called ionized Reynolds layer (Hoyle \& Ellis 1963; Reynolds 
1989). Using the available data at the Solar neighborhood, Boulares \& Cox 
(1990; hereafter BC) have incorporated the distended gas components, along with 
the vertical gradient of the magnetic field, in a model for the thick disk of 
the Galaxy. They found that a time averaged hydrostatic support of the disk 
requires scaleheights of about 1 kpc for both the ionized Reynolds layer and 
the magnetic and cosmic ray pressures. These extended gas layers are not
unique to our Galaxy, and recent observations indicate that they are probably 
common in spirals. Examples of diffuse ionized gas in external galaxies, that 
seem to be analogous to the Reynolds layer in our Galaxy, are the prominent
$H_{\alpha}$ diffuse halos in NGC 891 and NGC 5775, and to a lesser extent in 
NGC 4302 (Rand 1995). In particular, NGC 891 reveals $H_{\alpha}$ emission 
extending up to 3 kpc above the midplane (Rand, Kulkarni \& Hester 1990), along 
with other extraplanar structures such as worms (Dettmar 1990) and dusty 
filaments (Howk \& Savage 1997). Also, the magnetic fields in edge-on galaxies
can be traced for several kpcs above the disk (Hummel \& Beck 1995), and dusty 
filaments extending for more than 2 kpc above the plane of the host galaxy have 
been observed in NGC 253 and NGC 7331 (Sofue 1987). Some of these features are 
expected from the action of radiation pressure on dusty HI clouds (Franco 
\etal 1991), and from the spiral density wave (Martos \& Cox 1998). 

Regardless of some poorly known ISM parameters (such as the distribution and
filling factor of the hot gas), the existence of extended gas layers (neutral
or ionized) might have far reaching consequences for the structure and overall 
stability of the gaseous disk. Any given parcel of gas located above the 
scaleheight of the thin disk weighs more than the same material placed within
this disk. The inclusion of an extended gas layer, then, results in a disk 
model with a total interstellar pressure that is higher than the previously 
assumed thin disk values (this is in line with recent UV studies that indicate 
that the thermal pressure near the Sun can be a factor of 8 higher than 
previous estimates; Bergh\"ofer \etal 1998). The non-thermal pressure gradients 
at high latitudes, along with the tension of magnetic lines, must play a 
crucial role in the overall support of these extended disk structures (BC; 
Zweibel 1995). The presence of a thick magnetic disk, then, should drastically 
alter the results obtained with purely hydrodynamic models, or models in which 
the gradients are concentrated in a thin disk with a scaleheight of about 150 
pc. 

In this paper, we present two dimensional simulations of HVCs colliding with 
a thick disk, including both the purely hydrodynamic and the ideal 
magnetohydrodynamic regimes. Magnetized models are built from an isothermal gas
distribution, in which magnetic support at high $z$-locations is crucial for 
the initial equilibrium state. The models have the field lines parallel to the 
galactic disk, but we consider two different line orientations: field lines 
lying in the plane of motion of the HVC, and field lines that are normal to 
this plane. The results with these orientations, along with those of the purely 
hydrodynamic cases, allow us to isolate the effects of field tension: for lines 
lying in the plane of motion, magnetic tension reverses the motion of HVC 
material and creates an outflow at late times. For the case when the field 
lines are perpendicular to the plane of motion, which have no tension effects, 
the magnetic pressure prevents the cloud material from reaching deep into the 
disk. Thus, in either case, the magnetic field does not allow any mass exchange 
with the halo. In contrast, the non-magnetic cases (which demand a hot halo for 
the initial equilibrium) evolve without resistance and allow mass mixing. The
results for these non-magnetic models confirm previous results, but some new
features are uncovered here. Also, the resulting gas structures are now 
modified by the thicker and more pressurized nature of our models. The 
evolution of impinging HVCs, with a range of approaching angles and velocities, 
is followed with the MHD code ZEUS 3-D. The thick disk models are described in 
the next section. Section 3 deals with the numerical treatment of the problem. 
Section 4 presents our results, which are summarized and discussed in Section 
5.

\section{Magnetized Disk Models}

The magnetic disk model is plane-parallel. Two forms of pressure, thermal
and magnetic, provide the support of the initial magnetohydrostatic equilibrium
state against the gravitational field provided by the disk stars. Our model 
does not include cosmic-ray pressure. The density and gravitational 
acceleration functions are given by:  
\begin{eqnarray}
{\rho}(z) & = & \rho_0 [0.6e^{-\frac{z^2}{2(70pc)^2}}+
                 0.3e^{-\frac{z^2}{2(135 pc)^2}}
               + 0.07 e^{-\frac{z^2}{2(135 pc)^2}} \\ \nonumber
           &   & + 0.1e^{-\frac{|z|}{400 pc}}+0.03e^{-\frac{|z|}{900 pc}} ]
               \ \ \ {\rm cm}^{-3}
\end{eqnarray}
and
\begin{equation}
g(z) = 8\times 10^{-9}(1-.52e^{-\frac{|z|}{325 pc}}-.48e^{-\frac{|z|}{900 pc}}
)
       \ \ \ {\rm cm\, s}^{-2},
\end{equation}
\noindent
where the midplane gas density is $\rho_0 =2.24\times 10^{-24}$ g cm$^{-3}$.
This density distribution adequately describes the observed gas $z$-structure
in the solar vecinity, as discussed by BC. The functional form for the
gravitational acceleration $g$ is taken from Martos (1993), and provides a 
good fit to the data of Bienaym\' e, Robin \& Cr\' ez\' e (1987). 

The total pressure is given by $p(z) = -\int_z^{z_{ext}} \rho g \, dz$, with 
the boundary condition $p(z_{ext} = $5 kpc)=0, and is numerically solved and 
set equal to the sum of the thermal, $p_t=n(z)kT_{eff}$, and magnetic, $p_b =  
B^2(z)/8\pi $, terms. The midplane values are taken from BC: total pressure
$p(0) =2.7\times 10^{-12}$~dyn~cm$^{-2}$ (this is 20\% higher than the thermal
pressure value derived by Bergh\"ofer \etal 1998), a magnetic field strength of
$B(0) = 5\ \mu$G, and an effective disk temperature of $T_{eff}(0)=10900$ K. 
For simplicity, the magnetic model we adopt, which may be called a ``warm'' 
magnetic disk model, is defined by $T_{eff}(z) = T_{eff}(0)$ (independent of 
$z$). Thus, the implicit sound speed of this warm model is similar to the 
observed velocity dispersion of the main HI cloud component, $\sim 8$ km 
s$^{-1}$. The total magnetic field intensity at midplane adopted in this model 
includes contributions of the orderly (with a strength of $\sim 2 \ \mu$G) and 
the dominant random component values (see Heiles 1996). This results in a 
moderate field value for a spiral galaxy, $5\ \mu$G, because an average total 
field strength of 19 $\mu$G has been derived for the disk of NGC 2276 (Hummel 
\& Beck 1995). We assume magnetic field lines that are parallel to the 
midplane, as indicated by data near the plane (see Valle\'e 1997), in our 
initial magnetohydrostatic states. 

In the 2-D MHD regime, the adopted warm disk model is Parker unstable (Martos 
\& Cox 1994) and, from a linear stability analysis for the undular mode, we 
have found that the minimum growth time and the corresponding wavelength are 60 
Myr and 3 kpc, respectively (Kim \etal 1999). The 2-D models indicate that the 
density enhancements become clearly apparent on timescales of the order of 100 
Myr (Franco \etal 1995; Santill\'an \etal 1999). The instability can be 
triggered by a HVC collision (see Franco \etal 1995), but the HVC-disk 
interaction evolves in shorter timescales (the clouds are completely shocked in 
only a few Myr). Thus, given that the timescales for the two events are very 
different, here we do not discuss the appearance of the Parker instability, and 
a more complete description of the instability in a thick disk (\ie the linear 
analyses and the nonlinear evolution) will be reported elsewhere (Kim \etal 
1999; Santill\'{a}n \etal 1999). 

These magnetic models reflect conditions in which the total pressure decreases 
more slowly than the density as $z$ increases. At high $z$, the models mimic 
the expected dominance of non-thermal forms of pressure, and the effective 
signal speed is high. As a consequence, the compressibility of the plasma is 
effectively controlled by the magnetic term, and the medium is ''stiff`` and 
elastic (Martos \& Cox 1998). The thermodynamic regime of the runs, isothermal 
or adiabatic, can only alter that character to a certain extent, but the 
assumed magnetic field geometry will certainly affect the response to any 
interaction (further details of the properties of this thick, magnetic model 
can be found in Martos 1993 and Martos \& Cox 1994). The distinction between 
models with field lines in the plane of motion and perpendicular to the plane 
of motion, then, corresponds to whether the magnetic tension influences the 
dynamical evolution or not, respectively. Both types of magnetized models are 
inititated in equilibrium and with the same $z$-distribution for the total 
pressure. The total pressure distribution is shown in Figure 1a. The 
corresponding $z$-variations for the Alfv\'en and maximum magnetosonic (the
quadratic sum of the Alfv\'en and sound velocities) wave speeds are shown in 
Figure 1b. The magnetosonic speed is the effective signal speed for 
compressional waves, and has a rapid increase inside the disk but varies 
slowly, from 50 to 60 km s$^{-1}$, in a wide $z$ interval from 500 to 1500 pc.

For completness, we have considered an additional third model representing a
non-magnetic thick disk. This model maintains the same total pressure as in the
previous magnetic models. Hence, hydrostatic equilibrium determines the 
temperature structure along the $z$-axis, $T(z)= \int_{z_{ext}}^z \rho g \, dz
/kn(z)$, with the density and gravity distributions described above. The
resulting temperature and sound speed distributions are shown in Figure 1c.

\section{Numerical Method} 

The simulations are performed with the MHD code ZEUS-3D (version 4.2), which 
solves the three dimensional system of ideal MHD equations by finite 
differences on an Eulerian mesh (for a description of the code, see \cite{SNa}, 
1992b). The code can perform simulations in 3D but, due to computational 
restrictions, here we restrict the discussion only to two dimensional 
simulations. The effects of self-gravity and differential rotation of the 
Galaxy are not included in the present version. The role of self-gravity is not 
important at the densities considered here, but the effects due to the shear of 
the galactic disk may be important during the evolution. When differential 
rotation is included, there are at least two effects that may prove to be 
significant for the HVC-disk collisions: first, the shear can cause distortions
in the resulting gaseous structures (\eg Olano 1982; Pal\v{o}us, Franco \& 
Tenorio-Tagle 1990) and, second, it can trigger the appeareance of
magneto-rotational instabilities (\eg Chandrasekar 1960; Balbus \& Hawley 1991; 
Foglizzo \& Tagger 1994). 

In particular, the combined effects of the Parker and the magneto-rotational 
instabilities may lead to interesting results (Foglizzo \& Tagger 1994). These 
instabilities have different dynamical effects on the plasma and magnetic field 
lines. The Parker instability distort the magnetic lines, generating a vertical
component from an originally horizontal field and redistributing the gas in the
disk. The magneto-rotational process stretches the field lines radially and 
generates internal torques, driving radial gas flows. Depending on the initial 
state, the instabilities may interfere constructively, with the vertical field 
lines from the Parker mechanism feeding back onto the magneto-rotational 
mechanism. In other cases, however, both processes operate in a stabilizing 
manner (Foglizzo \& Tagger 1994). These issues are important and require 
detailed three dimensional studies with differential rotation that, 
unfortunately, are beyond our present capabilities. The 2-D results of the 
present paper cannot include the galactic shear, and the 2D computational 
domain lies in the plane defined by the azimuthal and vertical directions of 
the Galaxy.

Our frame of reference is one in which the Galactic gas is at rest, and the
origin of our 2D Cartesian grid is the local neighborhood. The coordinates 
($x$, $z$) represent distances along and perpendicular to the midplane, 
respectively. The $x$-axis is anchored at a constant galactocentric radius 
(\ie is quasi-azimuthal, defined by the tangent of the local field orientation
in our Galaxy; see Heiles 1994 and Valle\'e 1997), and the motion of the HVC
is always considered in the ($x$, $z$) plane. Two different setups for the
magnetic field are considered: $B$ parallel to the $x$-axis, in the plane of
motion of the HVC, and $B$ parallel to the $y$-axis, perpendicular to the plane
of motion (and no deformation of the field is introduced by the dynamics of the 
gas). The $y$-direction is also defined in the midplane, but it would
correspond to the galactocentric radial vector direction. 

For an efficient use of computer resources, we mostly worked with moderate 
resolutions of 200$\times$200 zones, but verified that the results were not 
different from those obtained in runs with resolutions of 400$\times$400 zones. 
We performed runs with a variety of different sizes but, for simplicity, the 
physical intervals of the simulations presented here are 3 kpc $\times$ 3 kpc 
(the $z$-axis runs from -1.5 kpc to 1.5 kpc). Thus, one zone has an extent of 
15 pc per dimension, or better, with our linear zoning scaling. The boundary 
conditions are cyclic (periodic) in $x$, and free outflow in $z$. The 
evolution was computed in both the isothermal ($\gamma = 1.01$) and adiabatic 
($\gamma = 1.67$) regimes, since explicit cooling or heating functions are not 
included in our numerical scheme. 
                        
For simplicity, all infalling clouds were given the same dimensions, $210\times 
105$ pc (longer in the $x$ direction), and they are threaded by the magnetic
field strength corresponding to their initial locations. We performed some runs
with other cloud 
sizes but, except for the sizes of the initial pertubations, the results are 
similar to the ones described here. Since the evolution is followed on the 
$x-z$ plane, and we have set the initial density of the clouds to 
$n$=1 cm$^{-3}$, the mass and energy densities of the models were $5.0\times 
10^{-25}$ g cm$^{-3}$ and $2.5\times 10^{-11}$ erg cm$^{-3}$ (this would 
correspond to a cloud mass and kinetic energy of $3.5 \times 10^5$ \MSUN \ and
$3.5\times 10^{52}\  v^2_{100}$ erg, respectively, where $v_{100}$ is the cloud 
velocity in units of 100 km s$^{-1}$, if we set the third dimension to the 
quadratic mean of the other two). 
 
We positioned the cloud centers at several selected heights, from 350 pc to 
4050 pc, and made a series of runs with different incoming velocities and 
incident angles. The velocity range spanned was from 0 to 200 km s$^{-1}$ (\ie 
from free-fall to nearly the largest observed approaching velocity), and the 
angles were varied from 0$^\circ$ to 60$^\circ$ with the vertical ($z$) axis. 
Regardless of the initial position of the cloud, the entire cloud is 
shocked in less than 3 Myr. The evolution of the interaction is fast and takes 
place in a relatively small region (with dimensions of several dozens of cell 
sizes). Thus, the details of the early shock structure (which depend on the 
initial cloud conditions) are not resolved in our simulations (the interested
reader can find a detailed discussion of high resolution simulations for cloud 
collisions in Klein \& McKee 1994 and Mac Low \etal 1994), and we focus here 
only in the larger scale outcome of the impact (\ie in structures of the order 
of a hundred pc or larger). A summary of the runs presented in this paper is 
given in Table 1.

\section{Results}

\subsection{Non--magnetic thick disk}

As stated before, previously published calculations of HVC impacts have been
performed with a thin Galactic disk model, and most of them are perpendicular
collisions in the purely hydrodynamic regime. We start by comparing them with 
our results obtained with the non-magnetic disk model described above. This 
non-magnetic model requires a hotter halo and the thermal sound speed increases 
rapidly inside the main disk (see Figure 1c).  

Loosely speaking, the basic structures formed by the collisions are similar to
those described in previous modeling. For instance, as found in earlier works,
the sizes and shapes of the shocked layers in the disk resemble some of the HI
supershells observed by Heiles (1984) in our Galaxy. There are, however, some
clear differences with previous results, and they are mainly due to our more 
extended gas distributions (\ie the structures formed at high $z$-locations are 
denser, better defined, and last longer than in the thin disk cases). Also, the 
resulting rear wakes are now completely formed and their morphologies and 
vorticities are clearly apparent. In particular, here we see one conspicuous 
structure, the tail, that has been either missed or disregarded in former 
studies (this is likey due to the fact that most previous models have located
the HVC much closer to the midplane). The importance of this feature is better 
appreciated in (magnetic and non-magnetic) oblique impacts, and may be one 
of the possible sources of turbulence in the Reynolds' layer (see Benjamin 1998 
and Tufte \etal 1998). 

\subsubsection{HVC with $V_{HVC}$ = 200 km s$^{-1}$, and $\theta = 0^\circ$}

Our first example is a purely hydrodynamic simulation of a collision 
perpendicular to the disk (impact angle $\theta = 0^\circ$). The evolution is 
shown in the four snapshots displayed in Figure 2. The simulation is performed 
in the isothermal mode and the HVC center position is located at 1250 pc from 
midplane, with an initial velocity of 200 km s$^{-1}$. In all the following 
figures, the density is shown in logarithmic grayscale plots and the velocity 
field is indicated by arrows sized proportionally to the local speed.
  
The first two snapshots show the initial conditions and the shock evolution at
3.2 Myr, respectively. The impact creates a strong galactic shock directed 
downwards, and a reverse shock that penetrates into the cloud. The galactic 
shock tends to move radially away from the location of impact, but momentum 
conservation keeps it strongest in the direction of motion of the impinging 
cloud. The lateral components of the shock, then, are milder and become a sonic 
perturbation in relatively short timescales (see Tenorio-Tagle \etal 1986 and
Franco \etal 1988). The cloud has been completely shocked at the time of the 
second snapshot, and the lateral shocks have already disappeared. The cloud 
mass is locked in the shocked layer and, due to its supersonic motion, a vacuum 
is formed behind the layer. This rear vacuum
begins to be filled up by material falling from higher locations, as well as 
from gas re-expanding from the shocked layer. This creates a pair of vortices, 
one at each side of the layer, and a plume, or tail, is formed at the central 
part of the rear wake. This is clearly seen in the third snapshot, at 9.5 Myr. 
At this time, the shocked layer is already collecting gas from the denser parts 
of the disk. The shock front decelerates as it penetrates into the disk, and 
reaches the midplane at about 13 Myr. After crossing the midplane, the shock 
accelerates in the decreasing density gradient, and blows out into the other 
side of the halo. The beginning of the blow out process is apparent in the last 
snapshot, at 19 Myr. The swirling motions of the rear wake, and the large 
extent (about 1 kpc) and shape of the shocked layer inside the disk are clearly 
displayed in this frame. A large fraction of the original cloud mass remains 
locked up in the shocked layer, and a small amount of it has re-expanded back 
into the rear wake and tail. In turn, the tail has expanded sideways and it has 
a density minimum at the central part. This minimum is promoted by the 
acceleration of the shock front after crossing the midplane. The size of the
perturbed region has grown close to 2 kpc in this last snapshot. 

\subsubsection{HVC with $V_{HVC}$ = 200 km s$^{-1}$, and $\theta = 30^\circ$}

Figure 3 shows the hydrodynamical evolution for an oblique collision at $\theta 
=30^\circ$. Again, the original cloud is located at 1250 pc from midplane, with 
200 km s$^{-1}$, and the run is performed in the isothermal mode. Now the cloud 
momentum has an important lateral component, which is conserved during the 
evolution because the gravitational force has only a $z$-component. The first 
two frames show the evolution of the shocked layer at 3.2 and 6.3 Myr, 
respectively. The initial cloud is again completely shocked in a relatively 
short timescale, before the first frame, and the vacuum left by the cloud 
motion is filled up by infalling material and by re-expansion of the shocked 
layer. The shapes of the interstellar structures are modified by the lateral 
velocity component, but the main features of the hydrodynamical evolution are similar to the ones described in the previous case. The motion of the shocked 
layer creates the rear wake (with vorticity and swirling motions), and a tail 
that extends downstream to locations close to the point of impact.

The tail is now denser and more conspicuous than in the previous case, and it 
has the appeareance of an elongated finger or cometary tail. There is a visible 
shock in the second frame (at 6.3 Myr), when the central tail is forming. The 
slower downward speeds, and the inclination of the structure, allows for the 
gas of the tail to catch up with the main body of the shocked layer. The 
velocity vectors within the structure are now larger, and clearly show the 
re-expansion into the rarefied regions.  The shear between this faster flow and
its surrounding medium is subject to Kelvin-Helmholtz instabilities (\eg Shore
1992), but we cannot resolve the instability here. The oscillatory motion of 
the finger-like tail, is due to the combined 
effects of the vorticity of the rear flow and the unresolved instability. The 
prominence of this tail structure increases with both increasing HVC velocities 
and larger collision angles.

The main shock crosses the midplane at about 13 Myr, and accelerates 
afterwards. As in the previous case, a density minimum is generated behind the 
accelerated layer. Given that the acceleration is promoted by the density 
gradient, is then directed along the $z$-axis and creates an elongation of the
structure in this direction. This is clearly apparent in the third frame, at 
22.2 Myr. Again, the shock front begins to blow out of the disk at about these
times. The tail has also re-expanded at this time, and its vorticity and 
oscillations have created a chaotic velocity field along the trajectory of the 
interaction.

The late times evolution displays a wealth of features, illustrated in the last
snapshot at 47.7 Myr (to provide a better perspective, the midplane is now 
located in the middle of the frame). The central parts of the disk are 
distorted and compressed, with complex structures extending towards the impact 
zone. The disk scaleheight is altered on both sides of the plane, and the 
interphase between the upper and lower disk layers is marked but wavy, as in a 
water-air interphase. At the incoming side, there are rounded tongues, and some 
elongated structures breaking out from the disk, accompanied by a pair of 
vortices above. At the other side, the blow out expansion is clearly apparent
and the front is reaching the lower edge of the grid.

\subsection{Magnetic disk with $B$ parallel to the $x$-axis: the role of 
tension}

The rigidity and elasticity given to the disk by the magnetic field is better 
accentuated in 2D when the plane of motion of the HVC is parallel to the field 
lines, and the colliding gas distorts the initial field configuration. We 
illustrate the response of these deformed field lines with three representative 
cases. In these cases, the tension of the magnetic field dominates the 
evolution, and the results are completely different from those of the purely
hydrodynamic cases. For the figures of these magnetic cases, where the 
densities and velocities are indicated as before, the $B$-field lines are now 
displayed with continuous lines. 

\subsubsection{HVC with $V_{HVC}$ = 200 km s$^{-1}$, and $\theta = 0^\circ$}

The first MHD simulation is illustrated in Figure 4. It corresponds to a 
collision perpendicular to the disk with cloud parameters identical to those of 
Figure 2 (cloud located at 1250 pc, with velocity 200~km~s$^{-1}$ and impact 
angle $\theta = 0^\circ$), except that the simulation is now performed in the 
adiabatic mode.  

The initial shock is strong and has a magnetic Mach number close to 4, with a 
compression factor approaching 3 (a very strong shock has a compression factor 
of 4, as in the non-magnetic case). With these parameters, there are thermal 
shocks on both sides of the main shock front at the early evolutionary stages. 
The lateral and downward shocks, however, disappear in relatively short timescales, and the MHD waves begin to move ahead of the shocked layer. The 
first snapshot shows the expansion at $t$ = 3.2 Myr after impact. A series of 
MHD waves are already driven in all directions, creating compressions and a 
round shell-like structure (the ``bubble'') ahead of the shocked layer. The 
lateral disturbances are Alfv\'en waves moving along the field lines, and the 
disturbances in the $z$-direction are magnetosonic waves that compress the 
field lines. The initial shock fronts (in either direction) move faster than 
any of these waves, and are responsible for the strong deformation of the 
initial field configuration but, as stated before, the key parameter 
determining the outcome of this interaction is the downward distorsion of the 
magnetic field. 

During the first 7 Myr, a substantial fraction of the energy goes into the 
compression and tension of the distorted field lines (the second frame shows
the evolution at 6.3 Myr).  Also, the disk material that is inside (and above) 
the distorted sections slides down along the field lines, like in an inclined 
plane, and settles down at the location of the magnetic valleys. Thus, the 
distortions disrupt the local hydrostatic equilibrium, and there is a clear 
infall of material in the perturbed region. This creates a dense ``head'' of 
the perturbation moving towards the disk. The fast magnetosonic wave, moving in 
the upper disk layers at an average speed of 50 km~s$^{-1}$, creates a strong 
perturbation as it enters into the denser parts of the disk (it becomes a very 
mild magnetic shock, that is apparent in the third frame, with a magnetic Mach 
number always close to unity) and crosses the midplane at $t \sim 12$ Myr. At 
about 8 Myr, the energy stored in the magnetic field begins to be released as 
the field lines rebound, reversing the motion of the gas and lifting the dense 
head (that has already penetrated a few hundred 
pc into the lower layers) back to the upper parts of the disk. Thus, as seen in 
the third and forth frames (9.5 and 15.9 Myr, respectively), a high-velocity 
outflow moving away from the disk is created. This is a novel result in which 
an incoming flow is forced to become an outflow by magnetic tension.

At about 13 Myr, and later on, the compressional and Alfv\'en waves carry most 
of the available energy. The magnetosonic waves are able to perturb the other 
side of the halo, and the Alfv\'en waves continue to drive the expanding 
structure and create gas infall outside the location of the bubble. Thus, the 
resulting structure is characterized by rising motions inside the bubble (from 
the tensil restoring motion) and by lateral infall outside it. 

Figure 5 shows a run with the same initial parameters of the previous model, 
but now the evolution is isothermal. In this case, as expected, the shocked gas
piles up in a thin dense shocked layer that carries the momentum of the cloud.  
Thus, now the compression and distortion of the field lines is more pronounced 
in the zones where the momentum of the dense shocked layer is 
concentrated, but the main evolutionary features are similar to the ones
described in the previous case. The evolutionary times shown in Figure 5 are
identical to those shown in Figure 4 (3.2, 6.3, 9.5, and 15.9 Myr). Comparison 
with Figure 4 illustrates that the deformation of the field lines is now more 
acute, leading to a sharp V-shaped form (almost a discontinuity) at 6.3 Myr. A 
thin, dense, vertical structure is formed from the material that slides down 
along the distorted field lines. The compressional wave also becomes a weak
MHD shock as it enters into the disk, and the lateral expansions, driven by Alfv\'en waves, are similar to the ones described in the previous case 
(outflows in the central zones, and inflows in the external regions).

Except for differences in details and timescales, cases initiated at other $z$ 
locations and with different velocities behave in similar ways as the ones 
described in Figures 4 and 5: the field prevents the penetration of the cloud 
material into the disk and creates a net outflow at the late times evolution. 
For instance, an isothermal case started at $ z =  350$ pc and with a speed of 
100~km s$^{-1}$ encounters a rapid tensional rebound at about $t = 6$ Myr. 

\subsubsection{HVC with $V_{HVC}$ = 200 km s$^{-1}$, and $\theta = 30^\circ$}

The oblique magnetic case is illustrated in Figure 6 by an isothermal collision 
with $\theta = 30^\circ$ and 200 km s$^{-1}$, as in Figure 3. The galactic 
shock has an important lateral component, producing a strong compression in the 
$x$-direction, as seen in the first and second frames (3.2 and 6.3 Myr, 
respectively). Once again, the magnetic field rebounds but the lines tend to 
recover their original configuration in shorter timescales than in the 
perpendicular cases. The motion of the shocked layer is again reversed as the 
lines rebound (third frame at 15.9 Myr), and a series of prominent disk 
oscillations and MHD waves are apparent during most of the evolution. The 
horizontal component of the flow is maintained for a longer time (for instance, 
it has a velocity of 47 km s$^{-1}$  at 20 Myr), and the patterns of the 
velocity fields and magnetic field distortions are completely different to 
those of the perpendicular cases. 

As before, the Alfv\'en waves detach from the shocked layer, and create a 
region with infalling gas that surrounds the shocked structure. The 
magnetosonic waves also traverse the disk (becoming a weak MHD shock), and 
perturb the other side of the halo. The tail is again formed behind the shocked
layer but the magnetic field now constraints the flows and 
quenches the vorticity. The dense head moves almost parallel to the plane after 
rebound. This creates a magnetic shear flow but the tension of the lines 
prevents the appeareance of Kelvin-Helmholtz instabilities (\eg Frank \etal 
1996; Malagoli \etal 1996; Jones \etal 1997).

The asymmetries in the distorted lines produce two important effects in the 
tail. First, the downstream (right) field distortion has a larger extent, with 
a softer slope, than the one created upstream (left). Thus, there is more mass
sliding down towards the tail from the downstream side. This provides
additional momentum to the tail, and creates a large rarified region behind it, 
that is maintained for a long timescale (up to the end of the run). Second, 
the gas that slides down from the upstream side provides an effective (ram 
pressure) force that opposes to the motion of the tail. This is a 
Rayleigh-Taylor unstable situation (\eg Shore 1992) but, as in the case of the 
Kelvin-Helmholtz instabilities, we cannot resolve the instability. The undular 
shape of the tail is long lived and is due to the unresolved instability. It 
is still present at the fourth frame, at 41.2 Myr (also note that the shape of
the field lines are already distorted due to the Parker instability). In 
summary, a complex network of asymmetrical features is 
created, but the distorted tail moving almost parallel to the field lines is 
the most prominent structure of the run. 

Runs with other approaching angles and velocities generate the same type of 
features, but with logical differences. As the incident angle is increased, 
the asymmetries are increased: the 
lateral component of the velocity is increased and the downward penetration is 
reduced, but the lateral effects become more pronounced. The amplitude of the 
maximum distortion in the field lines is reduced accordingly, and line 
rebouncing occurs at earlier evolutionary times. For angles larger than 
$\theta \sim 60^\circ$, the flow becomes 
almost parallel to the $x$-direction. Nonetheless, the lines oscillate due to 
the collision and the oscillations transmit MHD waves in all directions, 
creating perturbations that are weaker but similar to those described before.

\subsection{Cases with $B$ perpendicular to the $x-z$ plane: the absence of 
tension}

In theses cases, the magnetic field lines are oriented along the $y$-axis 
(pointing outside of the figures). The lines preserve their straight alignment 
in the $y$ direction as their footpoints are dragged along, in the $x-z$ plane,
by the flow motions. Thus, the lines are not distorted and there are no 
Alfv\'en waves in these case. Also, there 
are no tension effects, but the effects of magnetic pressure and field 
compression are certainly present. Thus, the cloud gas can travel longer paths 
and penetrate deeper into the disk. Also, the total pressure provides a very 
effective drag that slows down the flow faster than in the purely 
hydrodynamical case, and distorts the morphology of the shocked layer. The 
evolution of these runs is intermediate to the previous non-magnetic and 
magnetic-with-tension cases.

\subsubsection{HVC with $V_{HVC}$ = 200 km s$^{-1}$, and $\theta = 0^\circ$} 

Figure 7 shows a model with the same cloud parameters as those of Figures 2 and 
5 (isothermal evolution with 200 km s$^{-1}$ directed along the $z$-axis, and
the HVC is located at 1250 pc). The early times evolution resembles the one of 
the purely hydrodynamical case described in section 4.1.1. The shocked layer 
collects the gas and aquires a bow-shock form. The vacuum left behind the 
shocked layer creates a swirling gas inflow, and a pair of vortices are
apparent behind the shocked layer during most of the simulation. As before, a 
tail moving behind the shocked layer is created, and the differences with 
the non-magnetic case begin to be 
apparent when the decelerating shocked layer reaches the velocity of the 
effective signal speed (at about 2 Myr). After this moment, the precursor 
compressional perturbation begins to move ahead of the shocked layer (it is 
already apparent, and located some 100 pc ahead of the shocked layer, in the 
first frame at 3.2 Myr). As in the previous magnetic cases, the layer and its 
precursor wave sink into the disk (second frame at 6.3 Myr), and the precursor 
magnetosonic wave becomes a new magnetic shock wave that creates a second 
shocked layer. Due to the lack of tension, the new MHD shock in this case is 
stronger than in the previous magnetic cases (with a magnetic Mach number 
ranging increasing from 1 to almost 3, as it penetrates into the central 
regions).

The resulting new shocked layer collects only disk material, and is clearly 
present in the third frame at 9.5 Myr. This shocked disk layer is the one
that actually penetrates into the central parts of the disk, and excites the
excites the magnetosonic perturbation that goes into the other side of 
the halo (last frame at 19 Myr). The final fate of the first shocked layer, on 
the other hand, which is the one that contains the gas from the impinging 
cloud, is controlled by the magnetic pressure. The gas is forced to re-expand 
by the compressed magnetic lines that have had accumulated in the space between 
the two shocked layers, and it goes back to high $z$-locations. Again, then, 
the cloud material cannot penetrate into the disk.

\subsubsection{HVC with $V_{HVC}$ = 200 km s$^{-1}$, and $\theta = 30^\circ$}

Figure 8 shows a model with $\theta = 30^\circ$ and the same parameters as in
Figures 4 and 6. The evolution follows the same basic features that were described in these previous models: a wake with swirling motions and a tail are 
created behind the shocked layer, and a second shock front appears when the 
compressional wave begins to penetrate the denser parts of the disk (first and 
second frames at 3.2 and 9.5 Myr, respectively). Again, the structure and 
evolution of the tail plays an important role in the evolution. The gas of the 
tail catches up with the main body of the shocked layer, and a prominent and 
elongated, finger-like, flow structure is created. As in the previous 
non-magnetic case of Figure 4, the flow is subject to Kelvin-Helmholtz 
instabilities, and the oscillatory motion of the structure (clearly apparent in 
the third frame at 19 Myr) is due to vorticity and the unresolved 
instability. Similarly, the prominence of the finger-like structure increases 
with both increasing approaching velocities and larger collision angles.

In the case of the second shocked layer, there is one important difference with
respect to the previous perpendicular case: the two shocked layers are not aligned. The snapshot at 19 Myr shows that the new shock is directed along
the $z$-axis. This is due to the fact that the density gradient is directed
along this same axis. Thus, the second shocked layer moves perpendicular to the
plane of the disk, and the relative orientation between the two layers is 
sensitive to the approaching angle of the cloud. The strength of this second 
shock increases with increasing cloud velocities, but decreases with increasing 
collision angles. Again, as in all previous magnetic cases, the cloud material
is unable to penetrate into the disk.

The late times evolution of the tail is almost identical to the one of the
non-magnetic case in Figure 3. The rounded (almost circular) tongues and
vortices extending along the trajectory toward the impact region (last frame
at 41.2 Myr), are long lived (up to the end of the run).

\section{Discussion and Conclusions}

We have presented simulations of HVCs collisions, at different incidence angles 
and velocities, with magnetic and non-magnetic models of the Galactic thick 
disk. In general terms, the structures formed in the non-magnetic cases are 
similar to those discussed in previous studies, but some novel features are 
uncovered here: the motion of the shocked layer creates a rear wake, with 
vorticity, and a conspicuous tail. In oblique collisions, the tail becomes more 
prominent and aquires an oscillatory motion that leads to a highly chaotic,
turbulent, velocity field along the trajectory of the interaction. Also, in 
contrast with thin disk results, where the perturbed region has dimensions 
similar to those of the original HVC, the resulting structures are larger and 
better delineated.
 
The response of a magnetized thick disk, on the other hand, reveals new aspects
of the interaction. Such a disk, with a strong magnetic support at high-$z$, 
also has important consequences for processes such as SN and superbubble 
evolution (\eg Slavin \& Cox 1993; Tomisaka 1994, 1998). In contrast with 
non-magnetic HVC-disk interactions, the cloud now encounters substantial 
resistance through its evolution in the halo and cannot merge with the gaseous disk. The results with a magnetic field indicate that the perturbed volume is 
certainly much larger than that of the non-magnetic counterparts. Moreover, if 
the disk is Parker unstable, as it is the case of our warm 
magnetic model, the collisions are able to excite different oscillation modes 
in the disk and the halo, and do trigger the Parker instability (see Franco
\etal 1995 and Santill\'an \etal 1998). With a $B$-field in the $x$-direction, 
the MHD waves propagate in all directions but any gas flow towards the disk is 
drastically quenched. The tension effectively stops the shocked gas, and 
reverses the motion of the flow, preventing any penetration of the original 
HVC mass into the disk and creating gas motions, with velocities in the range
of 40 to 60 km s$^{-1}$, away from the disk. Thus, at least for this restricted 
field geometry, the magnetic field represents an effective shield that prevents 
any direct mass exchange between the halo and the disk. 

For a $B$-field perpendicular to the plane of motion, the strength of the shock
also decreases rapidly but the compressional waves now have a more direct 
effect on the central disk. The results for this field topology, which has
magnetic pressure but behaves in a tensionless manner, are intermediate to the
non-magnetic cases and the ones with magnetic tension. The shocked layer can
move deeper into the disk, but the buildup of magnetic pressure in the
compressed gas eventually stops the motion of the layer and forces its
re-expansion. The compressional waves, however, are transformed into a new
secondary shock front that penetrates into the disk. This creates a more 
complex double shocked layer structure that lasts over several million of 
years. The interaction of the new shock front with the inner disk layers alters 
visibly the structure of the disk at large scales. Here again, however, the 
cloud material cannot penetrate very deep into the disk.

The uniformity and symmetry of the disk and field modeled here are obvious 
idealizations. At high $z$, a likely vertical component of the field should 
modify the gas transport, and field line wandering would probably make the 
fluid more viscous than modeled here. Thus, one might suspect that our results
would be altered in a 3D simulation. For instance, as suggested by the referee,
the field lines in the disk would split apart and let the cloud to go through
with much less resistance. Thus, our conclusions about bouncing clouds cannot
be conclusive and this issue requires a 3D verification. Due to the stochastic 
nature of the Galactic magnetic field (Parker 1979), however, we anticipate 
that some aspects of the behavior found in the two magnetic field topologies 
considered here are bound to be present in more realistic 3-D cases. In 
particular, it is hard to imagine a situation in which a cloud would not have 
to fight magnetic tension from tangled and compressed field lines. Thus, at
least in some cases, a thick disk containing a bulky bundle of tangled lines could act as an effective shield against material penetration into the 
innermost layers of the disk. Also, as stated before, the combined effects of
the Parker and magneto-rotational instabilities require three dimensional
studies with differential rotation. We are already making test runs with 
additional field morphologies and, also, 3-D cases with a moderate resolution. 
The results are encouraging, and a detailed analysis will be presented 
elsewhere (Martos \etal 1999).

Summarizing, the magnetic field provides an adequate coupling for the energy 
and momentum exchange between the disk and the halo, but inhibits the mass
exchange. The interactions can create strong MHD perturbations, with a 
turbulent network of flows and vertical gas structures. Thus, the interstellar 
$B$-field topology plays a paramount role in the final outcome of the 
interaction with colliding clouds, and further studies with a magnetized disk will shed more light on the origin and fate of the HVC system.

\acknowledgments
We are grateful to Bruce Elemgreen, the referee, and Steve Shore, the editor,
for useful comments and suggestions. We also thank M. Norman, M. MacLow and R. 
Fielder for continued consultory on Zeus. JF acknowledges useful, and heated, 
discussions with Bob Benjamin and Bill Wall during the Interstellar 
Turbulence conference, in Puebla, Mexico. AS thanks Victor Godoy and Juan 
Lopez for their help with the visualization. This work has been partially 
supported by DGAPA-UNAM grant IN130698, CONACyT grants 400354-5-4843E and 
400354-5-0639PE, and by a R\&D grant from Cray Research Inc. The numerical 
calculations were performed using UNAM's ORIGIN-2000 supercomputer.

%---------------------------------------------------------------------

\clearpage

%\documentstyle[apjpt4]{article}
%\begin{document}
\begin{deluxetable}{lccccl}
%\scriptsize
%\footnotesize
%\small
\tablecolumns{6}
\tablecaption{Parameters of the 2-D and 2.5-D runs}
%\label{tbl-4}}
\tablewidth{0pt}
\tablehead{
\colhead{Case} & \colhead{Evolution} &
\colhead{$\vec B$} &
\colhead{$v_{\rm cloud}$} & \colhead{$\theta$} \\
\colhead{} & \colhead{} & \colhead{Direction} &
\colhead{(km s$^{-1}$)} & \colhead{(degrees)} 
}
\startdata
1 & Isothermal & - & 200 & 0  \nl
2 & Isothermal & - & 200 & 30 \nl
3 & Isothermal & x & 200 & 0  \nl
4 & Adiabatic  & x & 200 & 0  \nl
5 & Isothermal & x & 200 & 30 \nl
6 & Isothermal & y & 200 & 0  \nl
7 & Isothermal & y & 200 & 30 \nl

\enddata
\end{deluxetable}

%\end{document}

\clearpage

\begin{figure}
%\plotone{f1.ps}
\caption{ The frames show the $z$--distributions for: (a) the total pressure,
(b) the Alfv\'en and maximum magnetosonic speeds for the magnetic disk, and 
(c) the temperatures and sound speed for the non--magnetic disk.}
\label{fig_1}
\end{figure}

\begin{figure}
%\plotone{f2.ps}
\caption{Perpendicular collision (non--magnetic disk, and isothermal 
evolution). The sequence shows the density (gray logarithmic scale), and 
velocity field (indicated by arrows), at four selected times (0, 3.2, 9.5 and 
19 Myr). The maximum velocity values are 200, 166, 77 and 47 km s$^{-1}$, 
respectively. The midplane is located at $z=0$ kpc. The distance between tick 
marks in the frames is 500 pc.}
\label{fig_2}
\end{figure}

\begin{figure}
%\plotone{f3.ps}
\caption{Oblique collision (non--magnetic disk, and isothermal evolution). The 
sequence shows the density (gray logarithmic scale), and velocity field 
(arrows), at four selected times: 3.2, 6.3, 22.2 and 47.7 Myr. The maximum 
velocity values are 176, 113, 32 and 26 km s$^{-1}$, respectively. The midplane 
is located at $z=0$ kpc. The distance between tick marks is 500 pc.}
\label{fig_3}
\end{figure}

\begin{figure}
%\plotone{f4.ps}
\caption{Perpendicular collision (magnetic disk, $B$ parallel to the $x$--axis,
and adiabatic evolution). The sequence shows the density (gray logarithmic 
scale), velocity fields (arrows), and magnetic fields (lines), at four selected 
times: 3.2, 6.3, 9.5 and 15.9 Myr. The maximum 
velocity values are 128, 71, 84 and 77 km s$^{-1}$, respectively. The midplane 
is located at $z=0$ kpc. The distance between tick marks is 500 pc.}
\label{fig_4}
\end{figure}

\begin{figure}
%\plotone{f5.ps}
\caption{Perpendicular collision (magnetic disk, $B$ parallel to the $x$--axis, and isothermal evolution). The sequence shows the density (gray logarithmic 
scale), velocity field (arrows), and magnetic fields (lines), at four selected 
times: 3.2, 6.3, 9.5 and 15.9 Myr. The maximum velocity values are 137, 72, 95 
and 50 km s$^{-1}$, respectively. The midplane is located at $z=0$ kpc. The 
distance between tick marks is 500 pc.}
\label{fig_5}
\end{figure}

\begin{figure}
%\plotone{f6.ps}
\caption{Oblique collision (magnetic disk, $B$ parallel to the $x$--axis, and 
isothermal evolution). The sequence shows the density (gray logarithmic scale),
velocity field (arrows), and magnetic field (lines), at four selected times:
3.2, 6.3, 15.9 and 41.2 Myr. The maximum 
velocity values are 149, 82, 49 and 48 km s$^{-1}$, respectively. The midplane 
is located at $z=0$ kpc. The distance between tick marks is 500 pc.}
\label{fig_6}
\end{figure}

\begin{figure}
%\plotone{f7.ps}
\caption{Perpendicular collision (magnetic disk, $B$ parallel to the $y$--axis, 
and isothermal evolution). The sequence shows the density (gray logarithmic 
scale), and velocity field (arrows), at four selected times: 3.2, 
6.3, 9.5 and 19 Myr. The maximum velocity values are 160, 126, 96 and 78 km 
s$^{-1}$, respectively. The midplane is located at $z=0$ kpc. The distance 
between tick marks is 500 pc.}
\label{fig_7}
\end{figure}

\begin{figure}
%\plotone{f8.ps}
\caption{Oblique collision (magnetic disk, $B$ parallel to the $y$--axis, and 
isothermal evolution). The sequence shows the density (gray logarithmic scale), 
and velocity field (arrows), at four selected times: 3.2, 9.5, 19 
and 41.2 Myr. The maximum velocity values are 173, 80, 56 and 38 km s$^{-1}$, 
respectively. The midplane is located at $z=0$ kpc. The distance between tick 
marks is 500 pc.}
\label{fig_8}
\end{figure}

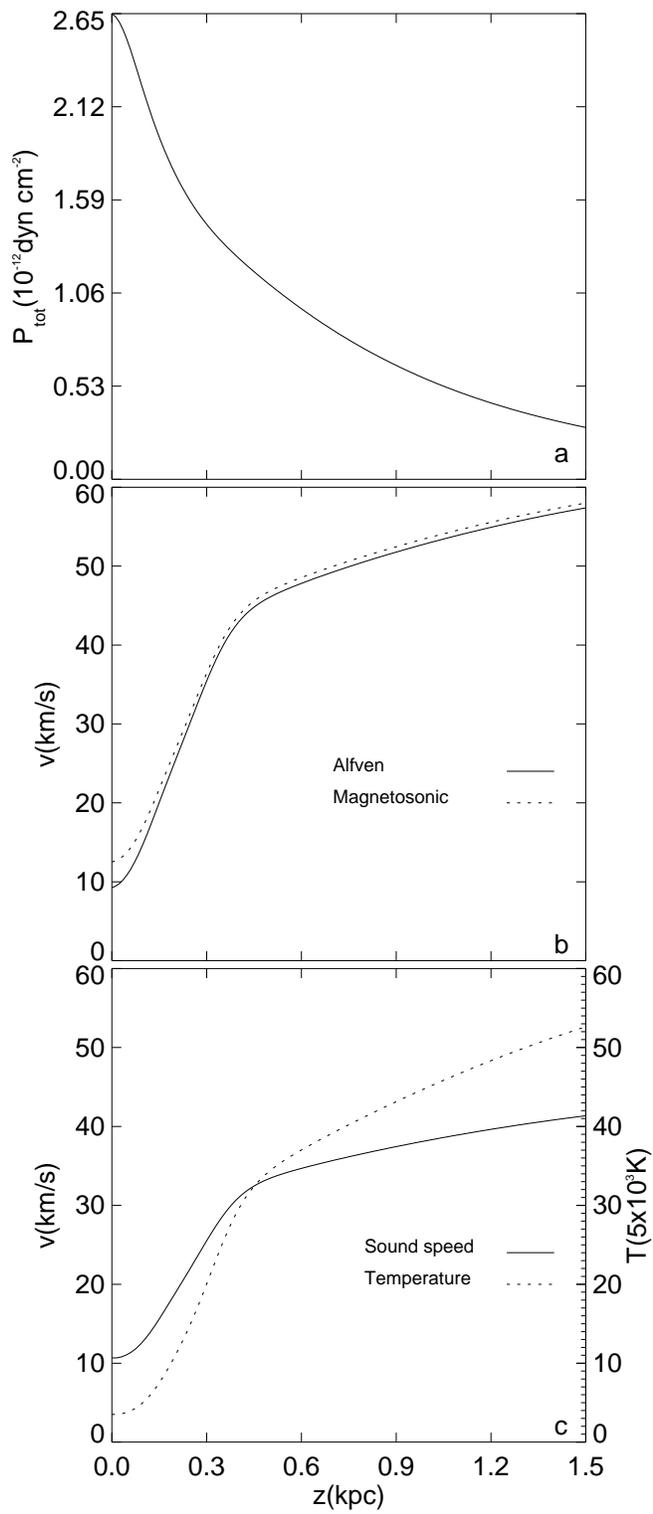
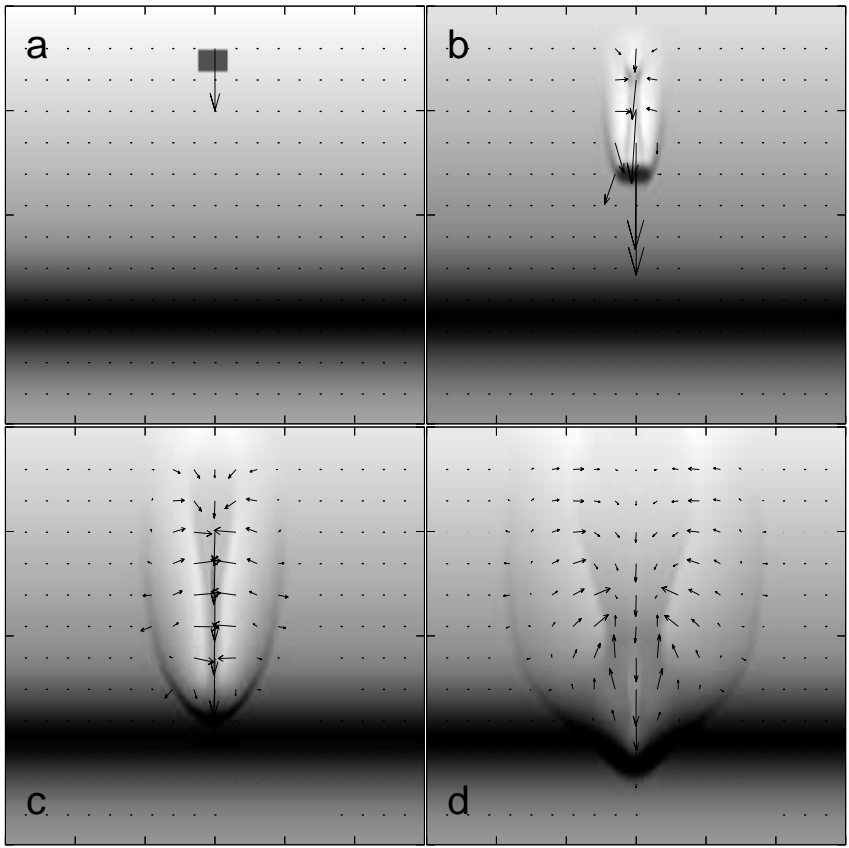
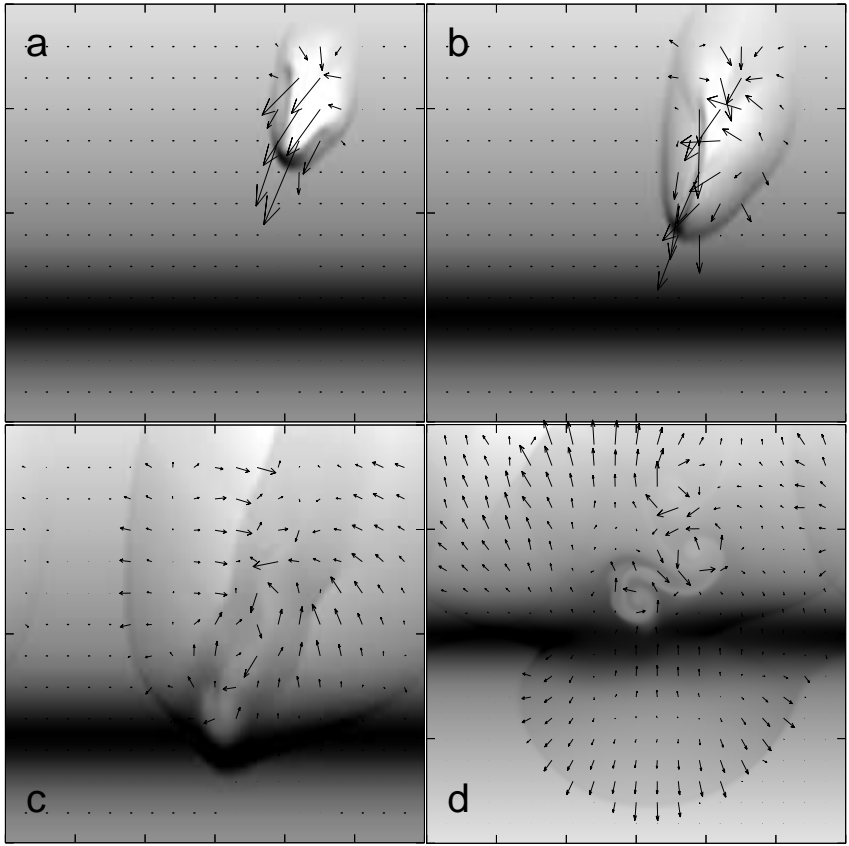
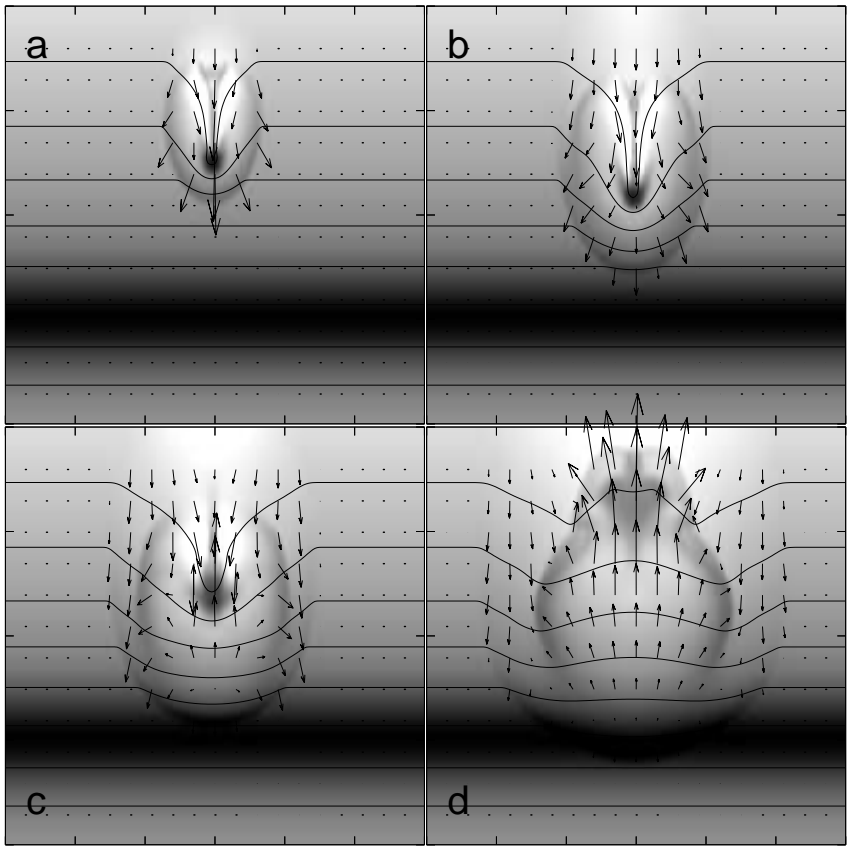
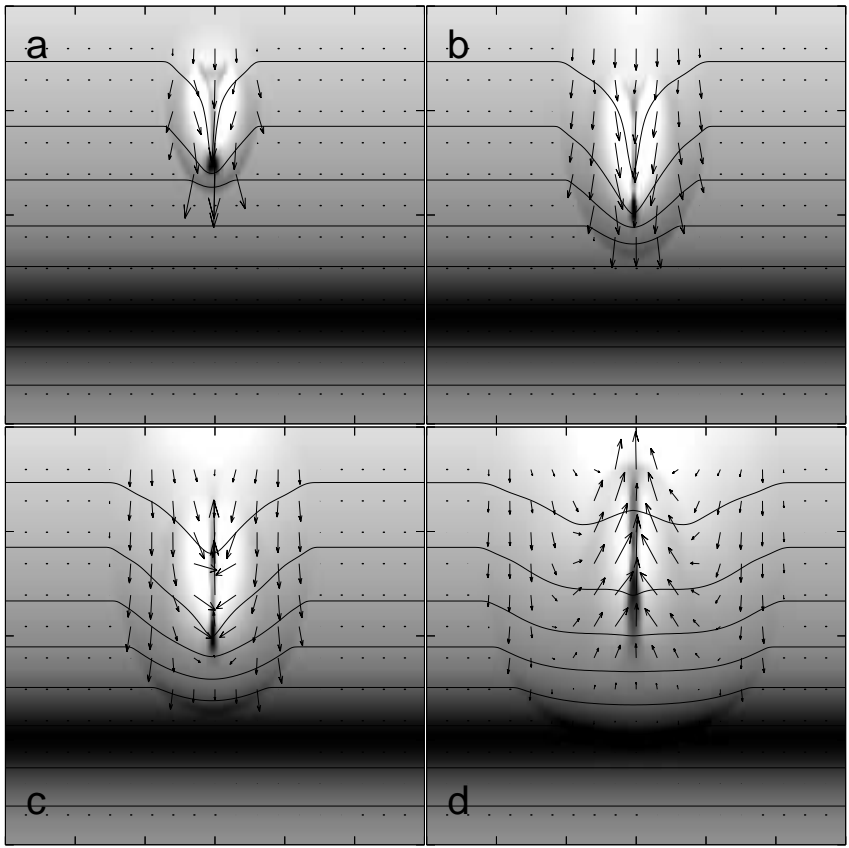
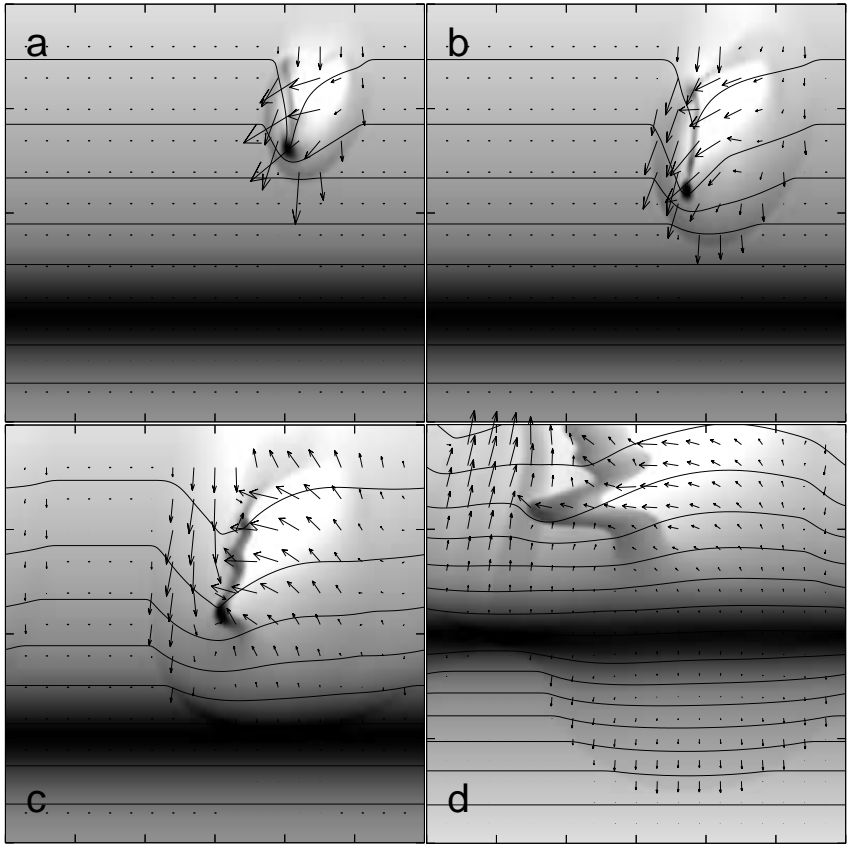
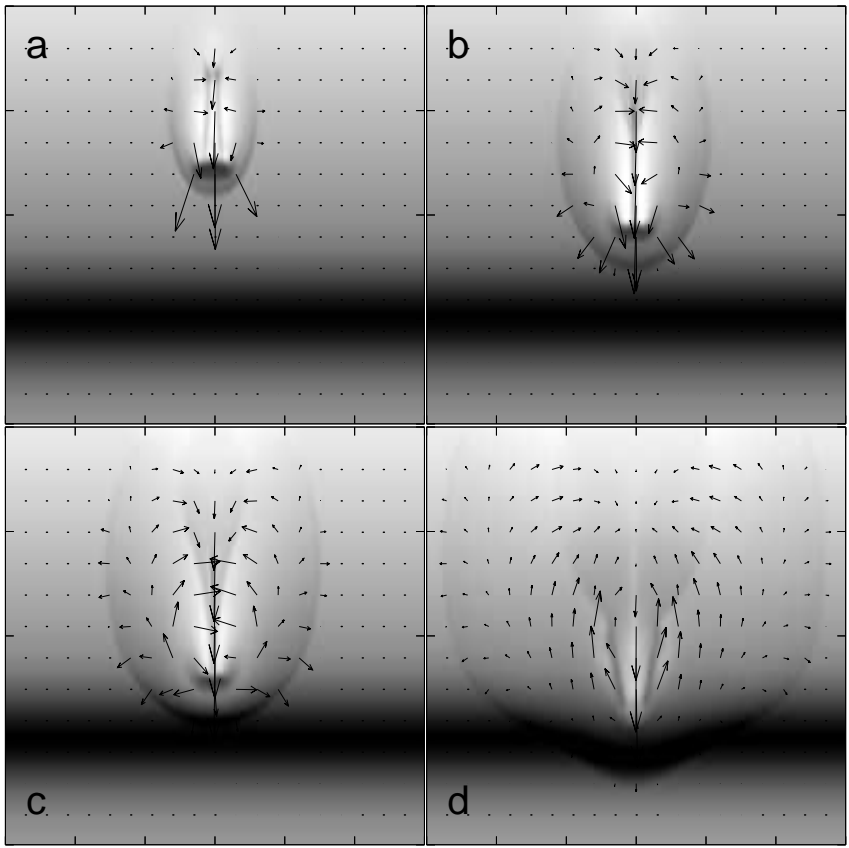
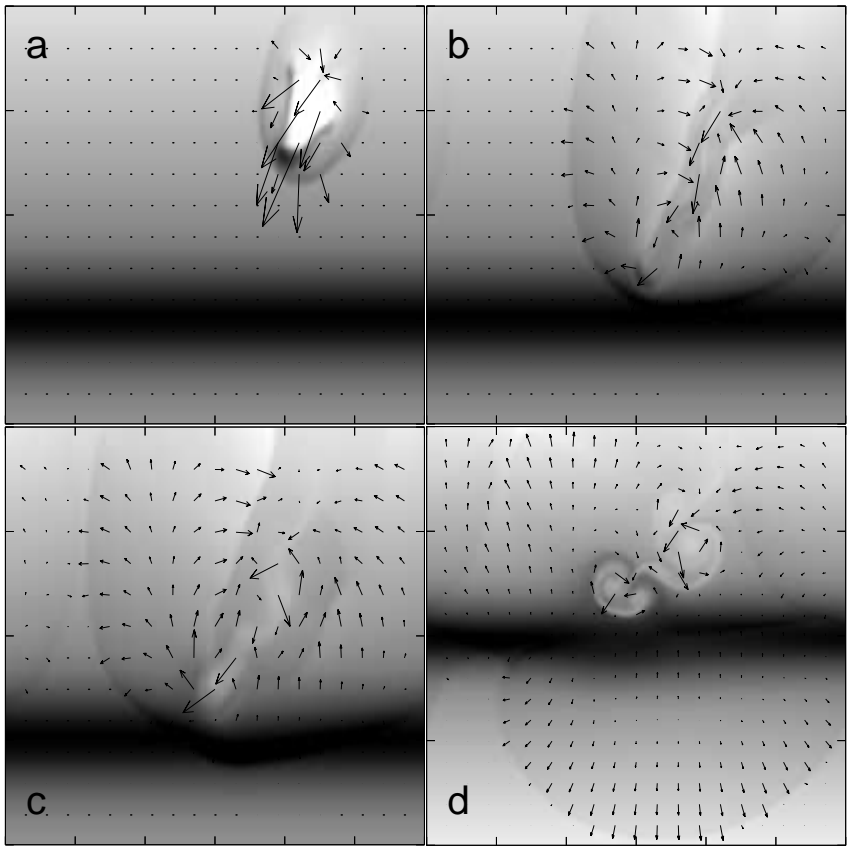
\end{document}